\documentclass[prb,twocolumn,tbtags,superscriptaddress,floatfix]{revtex4}
\usepackage{amssymb}
\usepackage{graphicx,amsmath}
\usepackage{bm}
\usepackage{times}

\begin{document}

\title{Amplification and attenuation of a probe signal by doubly-dressed
states}

\begin{abstract}
We analyze a system composed of a qubit coupled to the electromagnetic
fields in two high quality quantum oscillators. A particular realization of
such a system is the superconducting qubit coupled to a transmission-line
resonator driven by two signals with frequencies close to the resonator's
harmonics. This doubly-driven system can be described in terms of the
doubly-dressed qubit states. Our calculations demonstrate the possibility to
change the number of photons in the resonator and the transmission of the
fundamental-mode signal over a wide parameter range exploiting resonances
with the dressed qubit. Experiments show that in the case of high quality
resonators the dressed energy levels and corresponding resonance conditions
can be probed, even for high driving amplitudes. The interaction of the
qubit with photons of two harmonics can be used for the creation of quantum
amplifiers or attenuators.
\end{abstract}

\date{\today }
\author{S. N. Shevchenko\footnote{These authors contributed equally to this work}}
\affiliation{B. Verkin Institute for Low Temperature Physics and
Engineering, Kharkov, Ukraine} \affiliation{V. Karazin Kharkov
National University, Kharkov, Ukraine}
\author{G. Oelsner$^*$}
\affiliation{Leibniz Institute of Photonic Technology, Jena,
Germany}
\author{Ya. S. Greenberg}
\affiliation{Novosibirsk State Technical University, Novosibirsk, Russia}
\author{P. Macha}
\affiliation{Leibniz Institute of Photonic Technology, Jena, Germany}
\author{D. S. Karpov}
\affiliation{B. Verkin Institute for Low Temperature Physics and Engineering, Kharkov,
Ukraine}
\affiliation{V. Karazin Kharkov National University, Kharkov, Ukraine}
\author{M. Grajcar}
\affiliation{Department of Experimental Physics, Comenius University, Bratislava, Slovakia}
\author{U. H\"{u}bner}
\affiliation{Leibniz Institute of Photonic Technology, Jena, Germany}
\author{A. N. Omelyanchouk}
\affiliation{B. Verkin Institute for Low Temperature Physics and Engineering, Kharkov,
Ukraine}
\author{E.~Il'ichev}
\affiliation{Leibniz Institute of Photonic Technology, Jena, Germany}
\maketitle

\section{Introduction}

A number of experiments with strongly driven superconducting qubits have
been interpreted in terms of interference between multiple Landau-Zener
trajectories \cite{Oli} or multiphoton transitions \cite{Shnyr}. Several
years ago, Chalmers group demonstrated that the dynamics of a qubit in an
intense microwave field can be convincingly described by the dressed-state
model \cite{Wilson07}. Indeed, a system composed of qubits coupled to the
electromagnetic fields in resonators, represents a mesoscopic analogue of
atoms coupled to light fields in optical cavities, and similar effects have
been studied on atomic systems \cite{Mollow72, Wu77, Zakrzewski91}. Due to
this analogy the mathematical apparatus applied for describing the atomic
systems, can be adapted to the mesoscopic ones.

In this paper, we consider the situation of a qubit coupled to two driving
fields. Similar doubly-driven systems have already been used for the
two-tone spectroscopy \cite{Wallraff07} (also referred to as the pump-probe
technique \cite{Papageorge12, Silveri13}). Here, by driving the qubit with
one and probing the system at another (probe) frequency the energy level
structure can be reconstructed. It can be realized experimentally by making
use of two excited resonators \cite{Reuther10, Zhao13} or two modes of the
same resonator~[\onlinecite{Wilson07,Bishop09,Fink09,Hocke12}].

The notion of dressed states \cite{Coh-Tan, Nakamura01, Liu06, Sun11} can be
extended to two driving signals. In this case, one can describe the
aggregate system in terms of the doubly-dressed states. \cite{Yan01,
Alzar06, Saiko11} Note, that this approach is in contrast to the Floquet
method, in which the Floquet states make the quasi-classical counterpart of
the dressed states and their energies \cite{Silveri13, Satanin13, Abovyan13}.

In particular, due to the energy and information transfer between the
subsystems, the designated signal can be amplified or attenuated. \cite%
{Coh-Tan} Indeed, recently the amplification and attenuation of a probe
signal were studied in a number of experiments~[%
\onlinecite{Astafiev10,Oelsner12,Koshino13}]. Such processes can be
important for microwave engineering, including quantum amplifiers and
attenuators. In the work by Oelsner \textit{et al.} \cite{Oelsner12} such
kind of amplification was demonstrated. Here, the dynamics is as follows:
one mode induces Rabi oscillations in the qubit and, when the frequency of
these oscillations is adjusted to the second signal frequency, resonant
interaction can result in amplification or damping of the latter signal.
\cite{green02, Greenberg05, Greenberg07, Hauss08} For a practical
realization, the Rabi frequency is tuned into resonance with the oscillator
and thus provides a qubit-resonator energy exchange. Similar effects were
detected a long time ago \cite{Mefed} as NMR-amplification at the Rabi
frequency. Since the solid state quantum systems are scalable and tunable,
the effect of direct Rabi transitions can be used for microwave quantum
engineering.

The purpose of this article is to study in detail the doubly-driven qubit -
resonator system and to demonstrate its application to a realistic system.
To this end, we develop a theoretical formalism to describe the transmission
of a probe signal at the resonator's fundamental frequency. Our approach is
general and valid over a wide range of parameters. In particular, the
weak-probe limit can be relevant for probing the energy levels of the
qubit-resonator system. We not only discuss this case close to the Rabi
frequency, but also consider the resonant excitation of the coalesced system
for both strong driving and strong probing signals. This allows us to derive
analytic results for the qubit dressed, firstly, by the driving field and,
secondly, by the probing field. Importantly, our formalism explains
qualitatively and quantitatively the experimental results.

The rest of the paper is organized as follows. In Sec.~II the Hamiltonian of
the qubit interacting with two modes of the driven resonator is introduced
and the energy spectrum of the total system is calculated. The details of
those derivations are presented in the Appendix, where it is shown that the
aggregate system after transformations can be described as a dressed qubit.
The same transformations modify the master equation, which includes the
relaxation rates, as it is described in Sec.~III. The solution of the
Lindblad equation with the parameter-dependent relaxation rates is the
subject of Sec.~IV. In particular, it is demonstrated that the driving in
one harmonic influences the signal at other harmonics via the qubit. The
corresponding experimental results are presented in Sec.~V.

\section{Doubly-dressed states}

Consider a transmission-line $\lambda /2$-resonator with a single flux qubit
in it. The resonator is assumed to be driven by two signals: a low-amplitude
probing signal, with a frequency $\omega _{\mathrm{p}}$ close to the
resonator's fundamental frequency $\omega _{\mathrm{r}}$, and a
high-amplitude driving signal, with a frequency $\omega _{\mathrm{d}}$. The
driving frequency is considered to be close to the third harmonic frequency,
and hence one has to take into account this very harmonic component in the
Hamiltonian.

In the experimental realization \cite{Oelsner12, Oelsner10} the observables
relate to the fundamental mode and the system can be described by the
reduced Hamiltonian traced over the third-harmonic resonator mode. In the
Appendix we have rewritten the total system's Hamiltonian so that it
includes the strong driving signal as the renormalization of the qubit's
Hamiltonian, which can be interpreted in terms of the dressed states. There,
we start from the bare qubit, characterized by $H_{\mathrm{qb}}=-\frac{%
\Delta }{2}\tau _{x}-\frac{\varepsilon _{0}}{2}\tau _{z}$ in terms of the
Pauli matrices $\tau _{i}$; it has the bare energy level distance $\Delta E=%
\sqrt{\Delta ^{2}+\varepsilon _{0}^{2}}$. The qubit is considered to be
coupled to the two-mode resonator and the coupling characterized by the
value $\mathrm{g}_{1}$. So, the total Hamiltonian in the rotating-wave
approximation (RWA) is written as follows (see Eqs.~(\ref{H_qb-r}, \ref%
{Hqb+d}, \ref{Hqb})):
\begin{eqnarray}
\widetilde{H} &=&-\frac{\widetilde{\varepsilon }}{2}\sigma _{z}+\frac{%
\widetilde{\Delta }}{2}\sigma _{x}+\hbar \omega _{\mathrm{r}}a^{\dag
}a-\hbar \mathrm{g}_{\varepsilon }\left( a+a^{\dag }\right) \sigma _{z}+
\notag \\
&&+\xi _{\mathrm{p}}\left( ae^{i\omega _{\mathrm{p}}t}+a^{\dag }e^{-i\omega
_{\mathrm{p}}t}\right) .  \label{H_tilda}
\end{eqnarray}%
This describes the dressed qubit interacting with the resonator with the
renormalized coupling $\mathrm{g}_{\varepsilon }=\mathrm{g}_{1}\varepsilon
_{0}/\Delta E$ and probed by the signal with the amplitude $\xi _{\mathrm{p}%
} $. The qubit is now described by the Pauli matrices $\sigma _{i}$\ and the
resonator is described by the photon annihilation and creation operators, $a$
and $a^{\dag }$. The dressed bias $\widetilde{\varepsilon }$ and the
tunneling amplitude $\widetilde{\Delta }$ are defined by the driving
frequency $\omega _{\mathrm{d}}\ $and amplitude $A_{\mathrm{d}}$ either in
the weak-driving regime, at $A_{\mathrm{d}}<\hbar \omega _{\mathrm{d}}$,%
\begin{equation}
\widetilde{\varepsilon }=\Delta E-\hbar \omega _{\mathrm{d}},\widetilde{%
\Delta }=\Delta A_{\mathrm{d}}/2\Delta E,  \label{need_Ref0}
\end{equation}%
or in the strong-driving regime, where the energy bias is defined by the
detuning from the $k$-photon resonance, $\widetilde{\varepsilon }\rightarrow
\widetilde{\varepsilon }^{(k)}$, and the renormalized tunneling amplitude is
defined by the oscillating Bessel function, $\widetilde{\Delta }\rightarrow
\widetilde{\Delta }^{(k)}$, as follows%
\begin{equation}
\widetilde{\varepsilon }^{(k)}=\Delta E-k\hbar \omega _{\mathrm{d}},\text{ }%
\widetilde{\Delta }^{(k)}=\Delta \frac{k\hbar \omega _{\mathrm{d}}}{%
\varepsilon _{0}}J_{k}\left( \frac{A_{\mathrm{d}}}{\hbar \omega _{\mathrm{d}}%
}\frac{\varepsilon _{0}}{\Delta E}\right) .  \label{need_Ref}
\end{equation}%
These values define the dressed energy levels, $\widetilde{E}_{\pm }=\pm
\widetilde{\Delta E}/2$, where the distance between the energy levels is
\begin{equation}
\widetilde{\Delta E}=\sqrt{\widetilde{\varepsilon }^{2}+\widetilde{\Delta }%
^{2}},  \label{dressedDE}
\end{equation}%
which gives the dressed Rabi frequency $\Omega _{\mathrm{R}}=\widetilde{%
\Delta E}/\hbar $. The Hamiltonian (\ref{H_tilda}) directly brings us to the
problem of a qubit interacting with a weakly-driven fundamental-mode
resonator, e.g.~[\onlinecite{Om10}], with the following changes of notation
for the bias and gap in the Hamiltonian with the dressed ones: $\varepsilon
_{0}\rightarrow \widetilde{\varepsilon }$ and $\Delta \rightarrow \widetilde{%
\Delta }$.

Diagonalization of the time-independent part of the Hamiltonian (\ref%
{H_tilda}), i.e. at $\xi _{\mathrm{p}}=0$, gives the energy levels for the
system of the qubit coupled to two modes of the resonator - the
doubly-dressed states:%
\begin{eqnarray}
E_{\pm ,n} &=&\hbar \omega _{\mathrm{r}}(n+1)\pm \frac{\hbar \Omega _{n}}{2},%
\text{ }E_{\mathrm{gr}}=-\frac{\hbar \delta \widetilde{\omega }_{\mathrm{qb}}%
}{2},  \label{dds_energies} \\
\Omega _{n} &=&\sqrt{4\widetilde{\mathrm{g}}^{2}(n+1)+\delta \widetilde{%
\omega }_{\mathrm{qb}}^{2}},~~n=0,1,2,..., \\
\delta \widetilde{\omega }_{\mathrm{qb}} &=&\widetilde{\Delta E}/\hbar
-\omega _{\mathrm{p}}\text{, }\widetilde{\mathrm{g}}=\mathrm{g}_{1}\frac{%
\varepsilon _{0}}{\Delta E}\frac{\widetilde{\Delta }}{\widetilde{\Delta E}}%
\text{.}
\end{eqnarray}

The probing signal is described by the last term in Hamiltonian (\ref%
{H_tilda}). When the photon energy $\hbar \omega _{\mathrm{p}}$ equals the
energy difference $E_{\pm ,n}-E_{\mathrm{gr}}$, resonant energy exchange
between the probing signal and the doubly-dressed states takes place. This
resonant condition then reads
\begin{equation}
\omega _{\mathrm{r}}n\pm \sqrt{\widetilde{\mathrm{g}}^{2}(n+1)+\left( \frac{%
\delta \widetilde{\omega }_{\mathrm{qb}}}{2}\right) ^{2}}+\frac{\delta
\widetilde{\omega }_{\mathrm{qb}}}{2}=\omega _{\mathrm{p}}-\omega _{\mathrm{r%
}}.  \label{res_cond}
\end{equation}%
Whether the energy exchange bear an amplifying or attenuating character
depends on the relaxation parameters, which is the subject of the next
section.

In particular, at weak resonant probe ($\omega _{\mathrm{p}}=\omega _{%
\mathrm{r}}$)\ and when the coupling is negligible($\mathrm{g}_{1}\ll \omega
_{\mathrm{r}},\Delta $), from Eq.~(\ref{res_cond}) we obtain two sorts of
resonance conditions (in agreement with the results of the semi-classical
Floquet formalism \cite{Silveri13, Tuorila10}). The first one does not
include resonator excitation: $n=0$ and%
\begin{equation}
\delta \widetilde{\omega }_{\mathrm{qb}}=0.  \label{(i)}
\end{equation}%
This corresponds to direct exchange of excitation between the probing signal
and the dressed qubit: $\hbar \omega _{\mathrm{p}}=\widetilde{\Delta E}$. In
other words, the probing signal frequency matches the qubit's Rabi
frequency: $\omega _{\mathrm{p}}=\Omega _{\mathrm{R}}=\widetilde{\Delta E}%
/\hbar $. The second resonance condition involves the resonator excitation: $%
n=1$ and%
\begin{equation}
\delta \widetilde{\omega }_{\mathrm{qb}}=\omega _{\mathrm{p}}.  \label{(ii)}
\end{equation}%
This describes the two-photon process at $2\hbar \omega _{\mathrm{p}}=%
\widetilde{\Delta E}$.\cite{Hauss08}

We note that the resonant conditions (\ref{(i)}-\ref{(ii)}) can
alternatively be obtained by neglecting in Eq.~(\ref{H_tilda}) the
qubit-resonator interaction and the probing signal, $\mathrm{g}=\xi _{%
\mathrm{p}}=0$. Then we obtain the energy levels for the qubit-resonator
system $E_{\pm ,n}^{0}=\pm \widetilde{\Delta E}/2+\hbar \omega _{\mathrm{r}%
}n $. The resonant condition which involves the energy exchange between the
probing signal and the dressed qubit, $\hbar \omega _{\mathrm{p}%
}=E_{+,n}^{0}-E_{-,n^{\prime }}^{0}$, gives at weak resonant probing ($%
\omega _{\mathrm{p}}=\omega _{\mathrm{r}}$) two sorts of resonances. The
first one does not include resonator excitation: $n=n^{\prime }$. This
corresponds to Eq.~(\ref{(i)}). In other words, the probing signal frequency
matches the qubit's Rabi frequency: $\omega _{\mathrm{p}}=\Omega _{\mathrm{R}%
}=\widetilde{\Delta E}/\hbar $. The second sort of resonances involves the
resonator excitation: $n\neq n^{\prime }$. For example, for $n-n^{\prime }=1$
we have $2\hbar \omega _{\mathrm{p}}=\widetilde{\Delta E}$, which describes
the two-photon process.

\begin{figure}[htp]
\includegraphics[width=8.5cm]{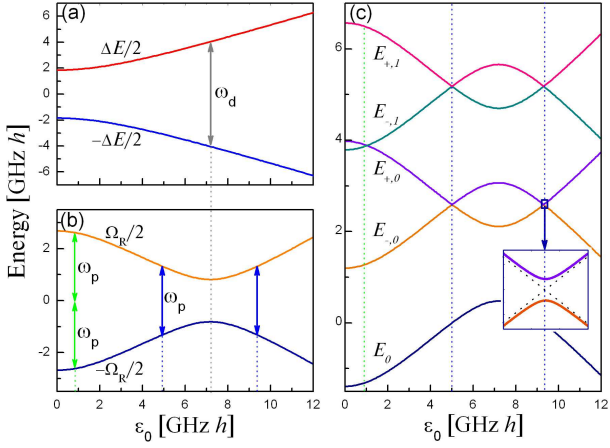}
\caption{(Color online) Bare, dressed, and doubly-dressed energy levels. The
bare qubit's energy levels, $\pm \Delta E/2$, are shown in (a); when they
are matched by the driving frequency $\protect\omega _{\mathrm{d}}$, the
qubit is resonantly excited. (At higher values of the bias $\protect%
\varepsilon _{0}$, the bare-qubit multiphoton excitation should be studied.)
The position of the resonance, $\hbar \protect\omega _{\mathrm{d}}=\Delta E$%
, is described by the avoided crossing of the dressed-state levels; the
dressed and averaged energy levels, $\pm \widetilde{\Delta E}/2=\pm \hbar
\Omega _{\mathrm{R}}/2$, are plotted in panel (b). When the dressed energy
levels are matched by the second (probe) signal, $\hbar \protect\omega _{%
\mathrm{p}}=\widetilde{\Delta E}$, a resonance interaction of the coalesced
system is expected, Eq.~(\protect\ref{(i)}). Also a resonance condition is
given by the two-photon process (\protect\ref{(ii)}).\ This is visualized as
the avoided crossings of the doubly-dressed states, plotted in panel (c).}
\label{Fig:En_levels}
\end{figure}

For the following calculations we switch to the dressed-qubit eigenstates
and after another RWA related to the probing signal we obtain the
time-independent Hamiltonian in the form (as it is described in the
Appendix).
\begin{equation}
H_{\mathrm{RWA}}=\hbar \frac{\delta \widetilde{\omega }_{\mathrm{qb}}}{2}%
\widetilde{\sigma }_{z}+\hbar \delta \omega _{\mathrm{r}}a^{\dag }a+\hbar
\widetilde{\mathrm{g}}\left( a\widetilde{\sigma }^{\dag }\!+\!a^{\dag }%
\widetilde{\sigma}\right) +\xi _{\mathrm{p}} \! \left( a\!+\!a^{\dag
}\right) ,  \label{H_RWA}
\end{equation}%
where $\delta \omega _{\mathrm{r}}=\omega _{\mathrm{r}}-\omega _{\mathrm{p}}$
and $\widetilde{\sigma }=\frac{1}{2}(\widetilde{\sigma }_{x}-i\widetilde{%
\sigma }_{y})$ is the qubit lowering operator and here tilde relates to the
new basis.

Finally, in Fig.~\ref{Fig:En_levels} we illustrate the energy levels studied
in this section. The parameters for the graph were taken for the system of
Ref.~[\onlinecite{Oelsner12}]: $\Delta /h=3.7$ GHz, $\omega _{\mathrm{r}%
}/2\pi =2.59$ GHz and $\omega _{\text{\textrm{d}}}/2\pi =3\omega _{\mathrm{r}%
}/2\pi =7.77$ GHz, $A_{\mathrm{d}}/h=7$ GHz, $\mathrm{g}_{1}/2\pi =0.8$ MHz,
and $\omega _{\mathrm{p}}=\omega _{\mathrm{r}}$. Figure~\ref{Fig:En_levels}
can be seen as the graphical description of dressing the dressed qubit,
which can be considered as the mesoscopic tunable analogue of the atomic
systems, as in Ref.~[\onlinecite{Greentree99}]. We also note that for the
small value of the coupling taken ($\mathrm{g}_{1}\ll \omega _{\mathrm{r}%
},\Delta $), the splitting (which defines the difference between the dressed
and doubly-dressed levels) is small, as shown in the inset. The solid lines
describe the doubly-dressed levels, while the dotted lines are for the
dressed levels, $\pm \widetilde{\Delta E}/2+n\hbar \omega _{\mathrm{r}}$.

\section{"Dressed relaxation"}

The Lindblad equation for the system can be written in terms of the bare
qubit and resonator states as follows~[\onlinecite{ScullyZubairy}]

\begin{equation}
\dot{\rho}=-\frac{i}{\hbar }[H,\rho ]+\sum\nolimits_{\alpha }\mathcal{L}%
_{\alpha }[\rho ],  \label{Lindbl_eq}
\end{equation}%
where the index $\alpha $ numerates different relaxation channels and the
respective Lindbladian superoperators are given by the formulas%
\begin{eqnarray}
\mathcal{L}_{\kappa } &=&\kappa \left( a\rho a^{\dag }-\frac{1}{2}\left\{
a^{\dag }a,\rho \right\} \right) , \\
\mathcal{L}_{\downarrow } &=&\Gamma _{1}\left( \sigma \rho \sigma ^{\dag }-%
\frac{1}{2}\left\{ \sigma ^{\dag }\sigma ,\rho \right\} \right) , \\
\mathcal{L}_{\phi } &=&\frac{\Gamma _{\phi }}{2}\left( \sigma _{z}\rho
\sigma _{z}-\rho \right) .
\end{eqnarray}%
These could also be written in the unified form
\begin{eqnarray}
\mathcal{L}_{\alpha } &=&L_{\alpha }\rho L_{\alpha }^{\dag }-\frac{1}{2}%
\left\{ L_{\alpha }^{\dag }L_{\alpha },\rho \right\} =  \label{compact} \\
&=&\frac{1}{2}\left[ L_{\alpha }\rho ,L_{\alpha }^{\dag }\right] +\frac{1}{2}%
\left[ L_{\alpha },\rho L_{\alpha }^{\dag }\right] .  \notag
\end{eqnarray}%
with the operators $L_{\alpha }$ given by the following: $L_{\kappa }=\sqrt{%
\kappa }a$, $L_{\downarrow }=\sqrt{\Gamma _{1}}\sigma $, $L_{\phi }=\sqrt{%
\Gamma _{\phi }/2}\sigma _{z}$. Here $\kappa $ is the decay rate of the
photons in the resonator, $\Gamma _{1}$ and $\Gamma _{\phi }$ are the
qubit's relaxation and dephasing rates.

After the transformations described in the previous section we arrive at the
Lindblad equation in the form of Eq.~(\ref{Lindbl_eq}) with the dressed
Hamiltonian $H_{\mathrm{RWA}}$ and with the dressed relaxation terms: $%
\widetilde{\mathcal{L}}=\mathcal{L}_{\kappa }+\widetilde{\mathcal{L}}%
_{\downarrow }+\widetilde{\mathcal{L}}_{\uparrow }+\widetilde{\mathcal{L}}%
_{\phi },$%
\begin{eqnarray}
\widetilde{\mathcal{L}}_{\downarrow } &=&\widetilde{\Gamma _{\downarrow }}%
\left( \widetilde{\sigma }\widetilde{\rho }\widetilde{\sigma }^{\dag }-\frac{%
1}{2}\left\{ \widetilde{\sigma }^{\dag }\widetilde{\sigma },\widetilde{\rho }%
\right\} \right) ,  \label{Gamma_down} \\
\widetilde{\mathcal{L}}_{\uparrow } &=&\widetilde{\Gamma _{\uparrow }}\left(
\widetilde{\sigma }^{\dag }\widetilde{\rho }\widetilde{\sigma }-\frac{1}{2}%
\left\{ \widetilde{\sigma }\widetilde{\sigma }^{\dag },\widetilde{\rho }%
\right\} \right) ,  \label{Gamma_up} \\
\widetilde{\mathcal{L}}_{\phi } &=&\frac{\widetilde{\Gamma _{\phi }}}{2}%
\left( \widetilde{\sigma }_{z}\widetilde{\rho }\widetilde{\sigma }_{z}-%
\widetilde{\rho }\right) .  \label{Gamma_fi}
\end{eqnarray}%
[Namely, to obtain these formulas, we applied the transformation $\widetilde{%
S}$ (see Appendix) to the original Lindbladians and kept only the terms
which survive under the RWA; this means keeping only the terms, which
contain $\widehat{1}$, $\widetilde{\sigma }_{z}$, or the product of $%
\widetilde{\sigma }$\ and $\widetilde{\sigma }^{\dag }$.] Thus, the dressed
Lindbladian superoperators in the sum $\sum\nolimits_{\alpha }\widetilde{%
\mathcal{L}}_{\alpha }[\widetilde{\rho }]$\ can be written in the compact
form of Eq.~(\ref{compact}), where $L_{\kappa }=\sqrt{\kappa }a$, $%
L_{\downarrow }=\sqrt{\widetilde{\Gamma _{\downarrow }}}\widetilde{\sigma }$%
, $L_{\uparrow }=\sqrt{\widetilde{\Gamma _{\uparrow }}}\widetilde{\sigma }%
^{\dag }$, $L_{\phi }=\sqrt{\widetilde{\Gamma _{\phi }}/2}\widetilde{\sigma }%
_{z}$. Here the tunable dressed rates for relaxation, excitation and
dephasing \cite{Hauss08, Ian10} are given by the expressions%
\begin{eqnarray}
\widetilde{\Gamma }_{\downarrow /\uparrow } &=&\frac{\Gamma _{1}}{4}\left(
1\pm \frac{\widetilde{\varepsilon }}{\widetilde{\Delta E}}\right) ^{2}+\frac{%
\Gamma _{\phi }}{2}\frac{\widetilde{\Delta }^{2}}{\widetilde{\Delta E}^{2}},
\label{G_down} \\
\widetilde{\Gamma }_{\phi } &=&\frac{\Gamma _{1}}{2}\frac{\widetilde{\Delta }%
^{2}}{\widetilde{\Delta E}^{2}}+\Gamma _{\phi }\frac{\widetilde{\varepsilon }%
^{2}}{\widetilde{\Delta E}^{2}}.  \label{G_fi}
\end{eqnarray}%
The Lindbladian $\mathcal{L}_{\kappa }$ describes the relaxation in the
resonator, $\widetilde{\mathcal{L}}_{\downarrow /\uparrow }$ describe both,
the relaxation and excitation in the qubit, while $\widetilde{\mathcal{L}}%
_{\phi }$ relates to the dephasing. \cite{Hauss08} Note that these "dressed"
relaxation rates appear due to taking into account photons of the driving
field, but neglecting the weak probing signal.

From Eq.~(\ref{G_down}) one can estimate the interplay of the relaxation and
excitation in the driven qubit. We note that the difference between the
excitation and relaxation rates is
\begin{equation}
\widetilde{\Gamma }_{\uparrow }-\widetilde{\Gamma }_{\downarrow }=\Gamma _{1}%
\widetilde{\varepsilon }/\widetilde{\Delta E},  \label{difference}
\end{equation}%
and the rates are equal at $\widetilde{\varepsilon }=0$, while for the red
and blue detuning, at $\widetilde{\varepsilon }\gtrless 0$ we have $%
\widetilde{\Gamma }_{\uparrow }\gtrless \widetilde{\Gamma }_{\downarrow }$
and either excitation or relaxation dominates.

In the limit of strong driving, the qubit's Hamiltonian is given by Eq.~(\ref%
{H_qb_k}), which coincides in form with Eq.~(\ref{Hqb}). This allows us
immediately to obtain the dressed relaxation rates in the vicinity of the $k$%
-th resonance (where $\Delta E-k\hbar \omega _{\mathrm{d}}\ll \Delta E$) by
simply replacing $\widetilde{\varepsilon }\rightarrow \widetilde{\varepsilon
}^{(k)}$ and $\widetilde{\Delta }\rightarrow \widetilde{\Delta }^{(k)}$ in
the formulas (\ref{G_down}-\ref{G_fi}).

To understand the impact of the effective relaxation terms, we consider the
reduction of the Lindblad equation to the Bloch-type one, similar as in
Ref.~[\onlinecite{Wilson10}]. Therefore, we start with the Lindblad equation
for the density matrix of the dressed qubit (neglecting the resonator
fundamental mode): $\widetilde{\rho }=\sum\nolimits_{i,j=0,1}\rho
_{ij}\left\vert i\right\rangle \left\langle j\right\vert $. Then from Eqs.~(%
\ref{Lindbl_eq}, \ref{Gamma_down}-\ref{Gamma_fi}) we obtain for the free
qubit evolution (taking into account that $\rho _{00}=1-\rho _{11}$):%
\begin{eqnarray}
\dot{\rho}_{11} &=&-(\widetilde{\Gamma }_{\uparrow }+\widetilde{\Gamma }%
_{\downarrow })\rho _{11}+\widetilde{\Gamma }_{\uparrow }, \\
\dot{\rho}_{01} &=&-\widetilde{\Gamma }_{2}\rho _{01},\text{ \ }\widetilde{%
\Gamma }_{2}=\widetilde{\Gamma }_{\phi }+\frac{\widetilde{\Gamma }_{\uparrow
}+\widetilde{\Gamma }_{\downarrow }}{2}.  \notag
\end{eqnarray}%
From here, in particular, it follows that the equilibrium population of the
excited state is defined by the rate $\widetilde{\Gamma }_{\uparrow }$: $%
\rho _{11}^{eq.}=\widetilde{\Gamma }_{\uparrow }/(\widetilde{\Gamma }%
_{\uparrow }+\widetilde{\Gamma }_{\downarrow })$.

\section{Impact of the qubit's driving on the transmitted signal}

The dynamical behavior of the system is described by the Lindblad equation (%
\ref{Lindbl_eq}) in combination with the Hamiltonian (\ref{H_RWA}). The
stationary solution can be found by assuming $\dot{\rho}=0$. In the limit of
small driving amplitude, as e.g. in Ref.~[\onlinecite{Om10}], an analytic
solution is possible.

From Eq.~(\ref{Lindbl_eq}) we obtain the equation of motion for the
expectation value of any quantum operator $A$
\begin{equation}
\frac{d\langle A\rangle }{dt}=-\frac{i}{\hbar }\langle \lbrack A,H_{\mathrm{%
RWA}}]\rangle +Tr\left( A\mathcal{\widetilde{L}}\right) ,  \label{YG1}
\end{equation}%
where $\langle A\rangle =Tr(A\rho )$, $\langle \lbrack A,H]\rangle
=Tr([A,H]\rho )$, the trace is over all eigenstates of the system. In our
system, the trace in Eq.~(\ref{YG1}) is over the photon states of the
fundamental mode $|n\rangle $ and the two qubit states $|\pm \rangle $. For
the expectation values of the operators $a$, $\widetilde{\sigma }$, $%
\widetilde{\sigma }_{z}$, $n=a^{\dag }a$ we obtain the following system of
equations (also called Maxwell-Bloch equations):
\begin{eqnarray}
\frac{d\left\langle a\right\rangle }{dt} &=&-i\delta \omega _{\mathrm{r}%
}^{\prime }\left\langle a\right\rangle -i\widetilde{\mathrm{g}}\left\langle
\widetilde{\sigma }\right\rangle -i\frac{\xi _{\mathrm{p}}}{\hbar },
\label{i} \\
\frac{d\left\langle \widetilde{\sigma }\right\rangle }{dt} &=&-i\delta
\widetilde{\omega }_{\mathrm{qb}}^{\prime }\left\langle \widetilde{\sigma }%
\right\rangle +i\widetilde{\mathrm{g}}\left\langle a\widetilde{\sigma }%
_{z}\right\rangle ,  \label{ii} \\
\frac{d\left\langle \widetilde{\sigma }_{z}\right\rangle }{dt} &=&-i2%
\widetilde{\mathrm{g}}\left( \left\langle a\widetilde{\sigma }^{\dag
}\right\rangle -\left\langle a^{\dag }\widetilde{\sigma }\right\rangle
\right) -\widetilde{\Gamma }_{+}\left\langle \widetilde{\sigma }%
_{z}\right\rangle -\widetilde{\Gamma }_{-},  \label{iii}
\end{eqnarray}

\begin{equation}
\frac{d\langle a\widetilde{\sigma }^{\dag }\rangle }{dt}=i(\delta \omega
+iG)\langle a\widetilde{\sigma }^{\dag }\rangle -i\widetilde{\mathrm{g}}%
(\langle \widetilde{\sigma }^{\dag }\widetilde{\sigma }\rangle +\langle
a^{\dag }a\widetilde{\sigma }_{z}\rangle )-i\frac{\xi _{\mathrm{p}}}{\hbar }%
\widetilde{\sigma }^{\dag },  \label{iv}
\end{equation}

\begin{eqnarray}
\frac{{d\left\langle {a{\widetilde{\sigma }_{z}}}\right\rangle }}{{dt}}
&=&-\left( {i\delta {\omega _{\mathrm{r}}}+Q}\right) \langle a\widetilde{%
\sigma }_{z}\rangle +i\widetilde{\mathrm{g}}\left\langle {\tilde{\sigma}}%
\right\rangle +{\widetilde{\Gamma }_{-}}\left\langle a\right\rangle
\label{asz} \\
&&-i\frac{{{\xi _{\mathrm{p}}}}}{\hbar }\left\langle {{\widetilde{\sigma }%
_{z}}}\right\rangle -2i\widetilde{\mathrm{g}}\left\langle {aa{{\tilde{\sigma}%
}^{\dag }}}\right\rangle +2i\widetilde{\mathrm{g}}\left\langle {{a^{\dag }}a%
\tilde{\sigma}}\right\rangle ,  \notag
\end{eqnarray}

\begin{equation}
\frac{d\langle n\rangle }{dt}=-\kappa \langle n\rangle +i\widetilde{\mathrm{g%
}}\left( \left\langle a\widetilde{\sigma }^{\dag }\right\rangle
-\left\langle a^{\dag }\widetilde{\sigma }\right\rangle \right) +\frac{i}{%
\hbar }\xi _{\mathrm{p}}(\langle a\rangle -\langle a^{\dag }\rangle ),
\label{YG2}
\end{equation}%
where $\widetilde{\Gamma }_{\pm }=\widetilde{\Gamma }_{\downarrow }\pm
\widetilde{\Gamma }_{\uparrow }$, $\delta \omega _{\mathrm{r}}^{\prime
}=\delta \omega _{\mathrm{r}}-i\kappa /2$, $\delta \widetilde{\omega }_{%
\mathrm{qb}}^{\prime }=\delta \widetilde{\omega }_{\mathrm{qb}}-i\widetilde{%
\Gamma }_{2}$, $\delta \omega =\omega _{\mathrm{qb}}-\omega _{\mathrm{r}}$, $%
G=\widetilde{\Gamma }_{2}+\kappa /2$, $Q=\kappa /2+\widetilde{\Gamma }_{+}$.
These equations present only the first few of the infinite ladder of coupled
equations, which include the higher-order correlations.

The stationary solution for this system can be found analytically if we
restrict the Hilbert space by $n=0,1$ states only, which is justified for
small amplitude probe signals (see, for example, Ref.~[\onlinecite{Om10}]).
However, in real experiments the number of fundamental photons is not so low
\cite{com}. In order to study this case we analyze below the full set of
equations (\ref{i}-\ref{YG2}).

By solving these equations we can obtain the transmission, as probed by a
network analyzer, which is defined in the way adopted in quantum optics \cite%
{Bishop09, Koshino13, Om10}
\begin{equation}
t=i\frac{\hbar \kappa }{2\xi _{\mathrm{p}}}\langle a\rangle .  \label{trans}
\end{equation}%
First we find the stationary solutions to the above equations for the case
when the interaction between the qubit and the resonator is absent, $%
\widetilde{\mathrm{g}}=0$. We obtain $\langle \widetilde{\sigma }\rangle
_{0}=0$,
\begin{equation}
\langle a\rangle _{0}=-\frac{\xi _{\mathrm{p}}}{\hbar \delta \omega
_{r}^{\prime }},  \label{YG4}
\end{equation}%
\begin{equation}
\left\langle \widetilde{\sigma }_{z}\right\rangle _{0}=-\frac{\widetilde{%
\Gamma }_{-}}{\widetilde{\Gamma }_{+}}.  \label{sz}
\end{equation}%
From Eq.~(\ref{YG4}) we obtain the transmission amplitude for the bare
resonator
\begin{equation}
|t_{0}|=\frac{\hbar \kappa }{2\xi _{\mathrm{p}}}|\langle a\rangle _{0}|=%
\frac{\kappa }{\sqrt{\kappa ^{2}+4\delta \omega _{\mathrm{r}}^{2}}},
\label{trans0}
\end{equation}%
and from Eq.~(\ref{YG2}) we obtain the average photon number
\begin{equation}
\langle n\rangle _{0}=-\frac{\xi _{\mathrm{p}}}{\hbar \kappa }2\text{Im}%
\langle a\rangle _{0}=\frac{4\xi _{\mathrm{p}}^{2}}{\hbar ^{2}(4\delta
\omega _{\mathrm{r}}^{2}+\kappa ^{2})}.  \label{n}
\end{equation}%
It follows that the condition for small $\langle n\rangle $ is $\xi _{%
\mathrm{p}}^{2}/\hbar ^{2}\kappa ^{2}\ll 1$.

In order to proceed further, we truncate the above infinite system of
equations. In the equations (\ref{iv}) and (\ref{asz}) we neglect the
correlations which contain three operators $\langle a^{\dag }a\widetilde{%
\sigma }_{z}\rangle $, $\langle aa\widetilde{\sigma }^{\dag }\rangle $, $%
\langle a^{\dag }a\widetilde{\sigma }\rangle $. This is justified if $%
\langle n\rangle $ is small. In terms of the density matrix this condition
is equivalent to $\langle 0\mid \rho \mid 0\rangle \gg \langle n\mid \rho
\mid n\rangle $, where $n\neq 0$. This means that the state $n=0$ is more
populated than other states with $n\neq 0$. For example, in Eq.~(\ref{iv})
the term $\langle a^{\dag }a\widetilde{\sigma }_{z}\rangle $ is neglected as
compared with $\langle \widetilde{\sigma }^{\dag }\widetilde{\sigma }\rangle
=(1+\langle \widetilde{\sigma }_{z}\rangle )/2$.

Next, we consider the qubit-resonator system in the absence of an external
probe, $\xi _{\mathrm{p}}=0$, $\widetilde{\mathrm{g}}\neq 0$. Then from the
steady-state solution of Eqs.~(\ref{iii}) and (\ref{iv}) we find the
steady-state polarization of the qubit in the resonator:
\begin{equation}
\langle \widetilde{\sigma }_{z}\rangle _{0}=-\frac{\widetilde{\Gamma }_{-}+A%
}{\widetilde{\Gamma }_{+}+A},  \label{sz2}
\end{equation}%
where
\begin{equation}
A=\frac{2\widetilde{\mathrm{g}}^{2}G}{(\delta \omega )^{2}+G^{2}},
\label{A=}
\end{equation}%
and the average photon number in fundamental mode:
\begin{equation}
\langle n\rangle _{\xi _{\mathrm{p}}=0}=\frac{1}{\kappa }\frac{A\widetilde{%
\Gamma }_{\upharpoonleft }}{\widetilde{\Gamma }_{+}+A}.  \label{nfm}
\end{equation}

Finally, we analyze the full set of truncated equations (\ref{i}-\ref{YG2}).
Considering the probing signal $\xi _{\mathrm{p}}$ to be weak, the qubit
operators acquire only small corrections: $\langle \widetilde{\sigma }%
_{z}\rangle =\langle \widetilde{\sigma }_{z}\rangle _{0}+O(\xi _{\mathrm{p}%
}^{2})$, $\langle \widetilde{\sigma }\rangle =O(\xi _{\mathrm{p}})$. Hence,
keeping the first-order approximation in $\xi _{\mathrm{p}}$ we obtain from
equations (\ref{i}), (\ref{ii}) and (\ref{asz}) the stationary solutions for
the intracavity field

\begin{equation}
\left\langle a\right\rangle =-\frac{\xi _{\mathrm{p}}}{\hbar }\frac{\delta
\widetilde{\omega }_{\mathrm{qb}}^{\prime }}{S\widetilde{\mathrm{g}}%
^{2}+\delta \widetilde{\omega }_{\mathrm{qb}}^{\prime }\delta \omega _{%
\mathrm{r}}^{\prime }},  \label{a}
\end{equation}%
where
\begin{equation}
S=\frac{{\delta \widetilde{\omega }_{\mathrm{qb}}^{\prime }\left[ {{%
\widetilde{\Gamma }_{-}}+i{{\left\langle {{\widetilde{\sigma }_{z}}}%
\right\rangle }_{0}}}\delta \omega _{\mathrm{r}}^{\prime }\right] }}{%
\widetilde{\mathrm{g}}{{^{2}}\left( {1+{{\left\langle {{\widetilde{\sigma }%
_{z}}}\right\rangle }_{0}}}\right) +i\delta \widetilde{\omega }_{\mathrm{qb}%
}^{\prime }\left( \delta \omega _{\mathrm{r}}^{\prime }{-i{\widetilde{\Gamma
}_{+}}}\right) }}.  \label{S}
\end{equation}%
Equation (\ref{a}) together with Eq.~(\ref{trans}) defines the transmission
amplitude. In particular, when the excitation rate is disregarded, $%
\widetilde{\Gamma }_{\uparrow }=0$, we have $\langle \widetilde{\sigma }%
_{z}\rangle _{0}=-1$. If, in addition, the relaxation rates are small (${{%
\widetilde{\Gamma }}}_{-},{{\widetilde{\Gamma }}}_{+}\ll \delta \omega _{%
\mathrm{r}}^{\prime }$) we get $S=-1$ and recover from (\ref{a}) the result
of Ref.~[\onlinecite{Om10}] obtained in the single-excitation regime (in the
first approximation in the probing-signal amplitude $\xi _{\mathrm{p}}$).

From Eq.~(\ref{YG2}) the average number of the fundamental-mode photons can
be found by:
\begin{eqnarray}
\langle n\rangle &=&-\frac{\xi _{\mathrm{p}}}{\hbar \kappa }2\text{Im}%
\langle a\rangle +\frac{i\widetilde{\mathrm{g}}}{\kappa }\left( \left\langle
a\widetilde{\sigma }^{\dag }\right\rangle -\left\langle a^{\dag }\widetilde{%
\sigma }\right\rangle \right) \\
&=&\frac{\xi _{\mathrm{p}}^{2}}{\hbar ^{2}\kappa }\frac{\kappa |\delta
\omega _{\mathrm{qb}}|^{2}+i\widetilde{\mathrm{g}}^{2}\left( S\delta \omega
_{\mathrm{qb}}^{\prime \ast }-S^{\ast }\delta \omega _{\mathrm{qb}}^{\prime
}\right) }{|S\widetilde{\mathrm{g}}^{2}+\delta \omega _{\mathrm{qb}}^{\prime
}\delta \omega _{\mathrm{r}}^{\prime }|^{2}}+\langle n\rangle _{\xi _{%
\mathrm{p}}=0}.  \notag
\end{eqnarray}

A solution can also be found in the limit of large excitation numbers, $%
\left\langle n\right\rangle \gg 1$, with a semi-classical approach.
Equations (\ref{i}-\ref{iii}) can be solved for an arbitrary value of $\xi _{%
\mathrm{p}}$, by removing the quantum correlations with the assumption that
the expectation values with qubit and photon operators to factorize: $%
\left\langle a\widetilde{\sigma }_{z}\right\rangle =\left\langle
a\right\rangle \left\langle \widetilde{\sigma }_{z}\right\rangle $, etc.
This is justified when the photon number is large, $\left\langle
n\right\rangle \gg 1$. \cite{Mu92, Ashhab09, Ella13} Then for the stationary
solutions, equations (\ref{i}-\ref{iii}) are equal to zero and the r.h.s.
can be rewritten in a more transparent form, excluding $\left\langle
\widetilde{\sigma }\right\rangle $,%
\begin{eqnarray}
\left\langle a\right\rangle &=&-\frac{\xi _{\mathrm{p}}}{\hbar }\frac{\delta
\widetilde{\omega }_{\mathrm{qb}}^{\prime }}{\left\langle \widetilde{\sigma }%
_{z}\right\rangle \widetilde{\mathrm{g}}^{2}+\delta \widetilde{\omega }_{%
\mathrm{qb}}^{\prime }\delta \omega _{\mathrm{r}}^{\prime }},  \label{A} \\
\widetilde{\Gamma }_{+}\left\langle \widetilde{\sigma }_{z}\right\rangle +%
\widetilde{\Gamma }_{-} &=&2i\frac{\xi _{\mathrm{p}}}{\hbar }\left(
\left\langle a\right\rangle -\left\langle a^{\dag }\right\rangle \right)
+2\kappa \left\langle a\right\rangle \left\langle a^{\dag }\right\rangle .
\label{B}
\end{eqnarray}%
Incidentally, the solution of the above equations gives, to first
approximation of the amplitude $\xi _{\mathrm{p}}$, the formulas (\ref{sz})
and (\ref{a}), being simplified to the result of Ref.~[\onlinecite{Om10}],
which is, strictly speaking, beyond the region of the validity of the
semi-classical approximation.

Note that the formula~(\ref{sz}) correctly describes the $k$-photon
excitations of the qubit (as e.g. in Refs.~[\onlinecite{SAN12, Li12}]),
\textit{i.e.} for small deviations $\widetilde{\varepsilon }^{(k)}=\Delta
E-k\hbar \omega _{\mathrm{d}}$ the stationary upper-level occupation
probability is described by Lorentzians positioned at $\Delta E(\varepsilon
_{0})=k\hbar \omega _{\mathrm{d}}$:%
\begin{equation}
\overline{P}_{\mathrm{up}}^{(k)}=\frac{1}{2}(1-\left\langle \widetilde{%
\sigma }_{z}\right\rangle )\approx \frac{1}{2}\frac{\widetilde{\Delta }%
^{(k)2}}{\frac{\Gamma _{1}}{\Gamma _{2}}\widetilde{\varepsilon }^{(k)2}+%
\widetilde{\Delta }^{(k)2}},
\end{equation}%
where $\Gamma _{2}=\Gamma _{\phi }+\Gamma _{1}/2$.

\section{Experiment}

\begin{figure}[tph]
\includegraphics{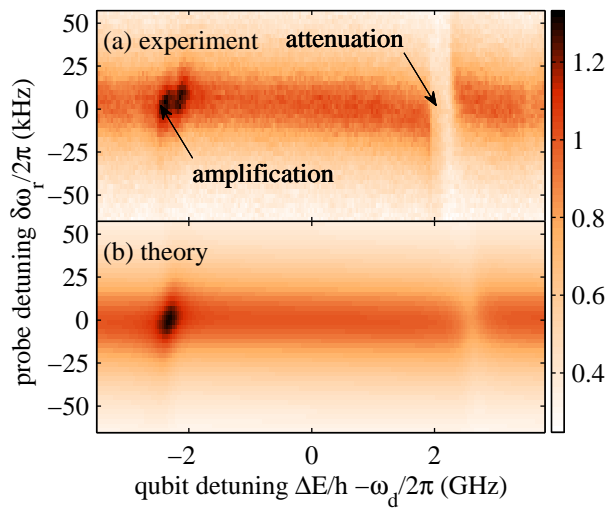}
\caption{(Color online) (a) Measured normalized transmission amplitude $%
\left\vert t\right\vert $ for the first qubit (see text) while a strong
driving signal is applied. The results are plotted in dependence on the
detuning between qubit frequency and driving signal, (controlled by the
qubit bias $\protect\varepsilon _{0}$), and the probing frequency detuning $%
\protect\delta \protect\omega _{\mathrm{r}}=\protect\omega _{\mathrm{p}}-%
\protect\omega _{\mathrm{r}}$. The amplification and the attenuation of the
transmitted signal is found in agreement with the resonance conditions, $%
\hbar \protect\omega _{\mathrm{p}}=\widetilde{\Delta E}$, Eq.~(\protect\ref%
{(i)}). (b) Normalized transmission calculated for the same parameters as in
(a) following Eq.~(\protect\ref{trans}) together with Eq.~(\protect\ref{a}).}
\label{Fig:expic1}
\end{figure}

To prove the theoretical description above, we fabricated a coplanar
waveguide resonator with resonance frequency of $\omega _{\mathrm{r}}/2\pi $
$=2.59$~GHz and a lossrate of $\kappa /2\pi =44$~kHz. Two qubit loops, with
sizes of $14\times7$~$\mu$m$^2$ for the first and $5\times5$~$\mu$ m$^2$ for
the second qubit are placed in the center of the resonator. With a high
quality factor of the resonator ($Q\approx $ $70000$) and rather weak
resonator-qubit couplings we ensure, that the decoherence of the coupled
system is not significantly increased in comparison to that of a bare qubit.
Indeed, we estimated the added width by the strong driving signal as $\sqrt{%
\left\langle N\right\rangle }\kappa $. So, even for photon numbers ranging
up to one million it is still in the same order as the qubits decoherence,
which is expected to be of the order of several 10 MHz. In that way the
definition of true energy states and corresponding resonance conditions is
justified. The sample was measured in a dilution refrigerator at a base
temperature of about $20$~mK. With independent measurements the minimal
level spacings $\Delta^{(1)} /h$ $=6$~GHz and $\Delta^{(2)} /h$ $=9.4$~GHz,
the persistent currents $I_{\mathrm{p}}^{(1)}$ $=60$~nA and $I_{\mathrm{p}%
}^{(2)}$ $=100$~nA, and the coupling constants $\mathrm{g}_{1}^{(1)}/2\pi $ $%
=8$~MHz and $\mathrm{g}_{1}^{(2)}/2\pi $ $=4$~MHz are defined for both
qubits (similar to Ref.~[\onlinecite{Oelsner10}]). Here superscripts (1) and
(2) denote the number of the qubit.

In a first set of experiments the bigger qubit was studied. A driving signal
at the fifth harmonic of the resonator $\omega _{\mathrm{d}}=5\omega _{%
\mathrm{r}}$ with a constant power was chosen. (We note that considering the
fifth harmonic instead of the third does not change any point in the
theoretical description of the system except of putting different values for
the driving frequency.) The transmission was measured at small detunings $%
\delta \omega _{\mathrm{r}} $ from the resonators fundamental mode, while
the energy bias $\varepsilon _{0}$ of the qubit was varied. The experimental
results are shown in Fig.~\ref{Fig:expic1}~(a) together with simulations
using Eq.~(\ref{trans}) Fig.~\ref{Fig:expic1}~(b). It can be seen that a
good agreement between the two was achieved for a relaxation rate $\Gamma _{%
\mathrm{1}}/2\pi =4$~MHz and a pure dephasing $\Gamma _{\phi }/2\pi =200$%
~MHz.

We find that if the qubit is detuned, the transmission amplitude is
independent of the bias $\varepsilon _{0}$ and is described by the
Lorentzian-shaped dependence on the frequency detuning, Eq.~(\ref{trans0}).
When the frequency of the probing signal is close to the Rabi frequency $%
\omega _{\mathrm{p}}\approx \Omega _{\mathrm{R}}=\widetilde{\Delta E}/\hbar $%
, a resonant energy exchange between the qubit and the fundamental mode of
the resonator results in amplification or attenuation of the transmitted
signal at red or blue detuning ($\widetilde{\varepsilon }\gtrless 0$)
correspondingly. The observation of such amplification was recently reported
by Oelsner \textit{et al.} in Ref.~[\onlinecite{Oelsner12}]. This can be
explained by the domination of relaxation or excitation, see Eqs.~(\ref%
{G_down}) and (\ref{difference}). Note that the inversion $\left\langle
\widetilde{\sigma }_{z}\right\rangle $, which defines the difference between
the occupation probabilities of the upper and lower qubit's levels, can be
estimated with Eq.~(\ref{sz}). This together with Eq.~(\ref{difference})
gives positive and negative inversion for $\widetilde{\varepsilon }>0$ and $%
\widetilde{\varepsilon }<0$, respectively. The effect of amplification and
attenuation of the transmitted signal can be related to the increase or
decrease of the cavity photon number. In this sense one can term these as
lasing and cooling of the resonator as in Ref.~[\onlinecite{Hauss08}].

\begin{widetext}

\begin{figure}[htp]
\includegraphics{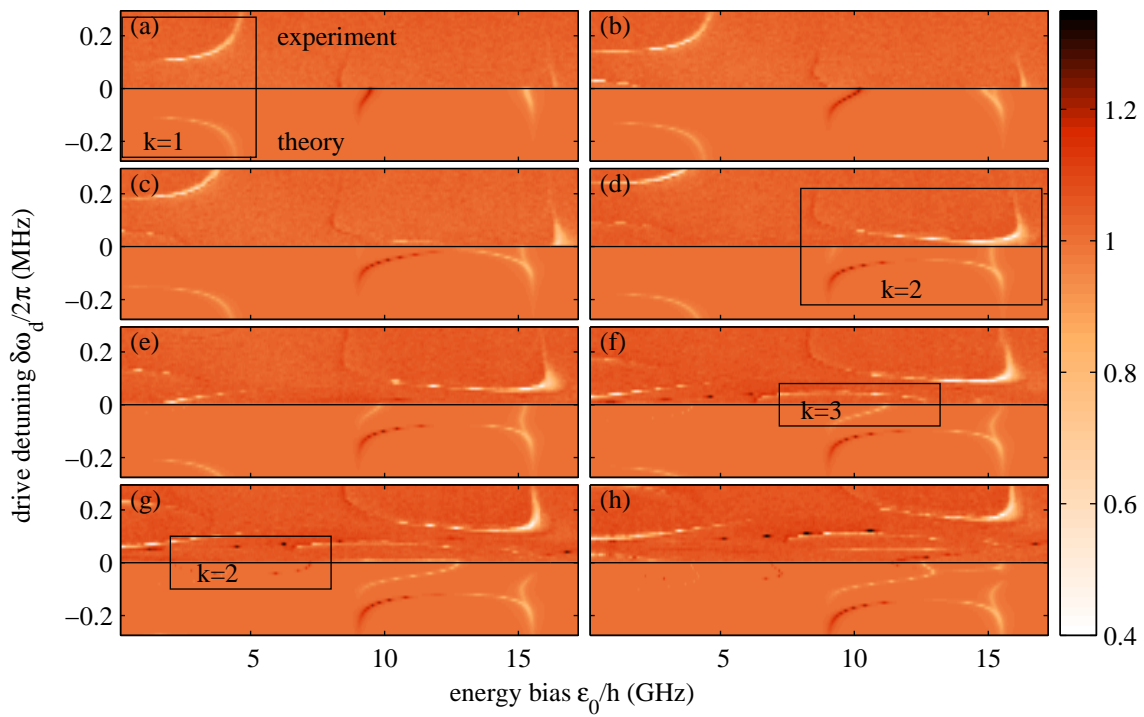}
\caption{(Color online) Normalized transmission amplitude through
the resonator at a probing frequency $\omega_\mathrm{p} = \omega_\mathrm{r}$ for different driving
amplitudes ranging from about $-131$~dBm in (a) to $%
-117$~dBm in (g) in $2$~dBm steps at the input of the resonator.
The transmission is plotted as a function of the energy bias
$\varepsilon_0$ of the second qubit and the driving frequency
detuning $\delta \omega _{\mathrm{d}} = \omega_\mathrm{d} - 3
\omega_\mathrm{r}$. The driving frequency is changed only around
the third harmonic in the order of its linewidth (about
$6\protect\kappa $). The presented results correspond to a
symmetric power dependence around the center frequency of the
third harmonic, since the resonator acts as bandpass filter. Each
plot is split into an experimental (positive detuning) and
a theoretical (negative detuning) part. For the calculations the
figures were split into regions for which Eq.~(\ref{trans}) was
used with a certain index $k$ in Eq.(\ref{need_Ref}). Several
features were highlighted with black rectangles to interrelate
theory and experiment (see text).} \label{Fig:expic2}

\end{figure}

\end{widetext}

In order to test our model in the strong driving regime, where Eq.~(\ref%
{need_Ref0}) is replaced by Eq.(\ref{need_Ref}), we analyze the response of
the system as a function of the driving amplitude. Here we consider the case
where the qubit gap is higher than the driving frequency. For this purpose
the smaller qubit was used. The transmission at the fundamental mode, $%
\omega _{\mathrm{p}}=\omega _{\mathrm{r}}$, was measured while changing the
frequency of the driving signal around the third harmonic frequency, and
consequently the driving amplitude at the qubit. The results are shown in
Fig.~\ref{Fig:expic2}, together with calculated data following from Eq.~(\ref%
{trans}). Several sharp lines of amplification (dark) and damping (light)
were experimentally observed. The number of lines increases with increasing
power and each of the lines corresponds to a resonance condition between the
dressed states and the probing signal. To understand their origin,
calculated transmission data was added into each plot. We split these
theoretical plots into regions and for each of them use Eq.~(\ref{trans})
with different index $k$ in Eq.~(\ref{need_Ref}). For high powers we note
that the levels of one step of the dressed ladder are equivalent to the ones
of the stairs above or below. In that way, we assume that Eq.~(\ref{trans})
is valid also for those regions where resonances between the lower level of
one stair and the higher level of the lower stair occur. To account for
these interaction we replace the splitting of the dressed states $\widetilde{%
\Delta E}^{(k)}$ by $\omega _{\mathrm{d}}-\widetilde{\Delta E}^{(k)}$. This
makes it possible to relate each of the resonance lines to one index $k$ and
to an interaction directly between the Rabi levels or levels of different
stairs.

We find a quantitative agreement between the theoretical predictions and the
experiment for a relaxation rate $\Gamma _{\mathrm{1}}^{(2)}/2\pi =6$~MHz
and a pure dephasing $\Gamma _{\phi }^{(2)}/2\pi =100$~MHz of the second
qubit. As examples we highlight several regions with black rectangles.
Direct resonances between the Rabi levels are marked in plot (a) for $k=1$
and in (d) for $k=2$. Note, that for $k=1$ no amplification is observed
since the driving frequency is chosen below the qubit gap. In Fig.~\ref%
{Fig:expic2}~(d) the power dependence of the resonance lines for $k=2$ show
a similar dependence as in Ref.~[\onlinecite{Hauss08}]. The marked regions
for damping in subplot (f) and amplification in (g) correspond to
interactions between levels of different stairs with the fundamental mode
and indices $k=3$ and $k=2$, respectively.

The representation of Fig.~\ref{Fig:expic2} gives the possibility to follow
the formation of the resonance lines and their change for increasing power.
Note, that the slight differences in the positions, which become more
pronounced for higher driving powers, are due to the limitation of our
model: we only consider two dressed levels. Therefore, the dynamics of the
system contains only the influence of the $k$-photon resonance, and the
theoretical predictions are close to the experiment where their influence
dominates. For instance, when different resonance lines come close to each
other, the two level approximation breaks down. For describing the full
dynamics it would be necessary to either include more levels of the dressed
ladder or to add corrections to the energy levels in one manifold (on one
stair of the dressed ladder) introduced by the stairs above and below.

Our experimental results show, that in the strong driving limit levels of
different photon numbers are strongly coupled and yield corrections to the
level splitting of the Rabi levels. Furthermore, due to the strong mixing of
the levels in the high driving case, resonance conditions can be defined
between each adjacent pair of levels.

\section{Conclusions}

We have considered the situation when a two-level system (qubit) is coupled
to two quantum oscillators. The emphasis was made on the specific
solid-state realization of a flux qubit coupled to a transmission-line
resonator, which is driven close to a harmonic and probed at the fundamental
frequency. The former driving signal is not directly observable and was
included into the considerations by the qubit's dressed states. Similarly,
the interaction of the dressed qubit with the resonator's fundamental mode
can be considered as the second dressing of the qubit's dressed states. When
the energy of the probing photons matches the dressed energy levels, a
resonant energy exchange results in either the amplification or attenuation
of the probing signal, depending on the tunable relaxation rates. We have
presented a detailed theoretical description of those processes, as well as
related the results to other studies, such as Refs.~[%
\onlinecite{Hauss08,
Wilson10, Silveri13}]. Our theoretical findings together with the
experimental results, presented here and in Ref.~[\onlinecite{Oelsner12}],
are useful for the description of the qubit-resonator systems in terms of
the dressed states. Especially since the high quality of the resonator
together with the rather weak coupling between the resonator and the qubit
prevent the energy levels from smearing out. In other words, although the
driving signal is strong, the added decoherence is still small, so that
sharp resonance conditions were observed. Furthermore, the idea of
transferring energy from one resonator's mode to another via a single qubit
may be useful for further applications.

\begin{acknowledgments}
This work was partly supported by NAS of Ukraine (Project No. 4/13 -NANO),
DKNII (Project No. M/231-2013), BMBF (UKR-2012-028).The research leading to
these results has received funding from the European Community's Seventh
Framework Programme (FP7/2007-2013) under grant agreement No. 270843.
Ya.S.G. acknowledges the partial support from the Russian Ministry of
Education and Science through the project TP 7.1667.2011 and from the German
Ministry of Science (BMBF) through the project RUS 10/015. M.G. was
supported by the Slovak Research and Development Agency under the contracts
APVV-0515-10 and DO7RP-0032-11. SNS thanks S. Ashhab for useful comments.
SNS and DSK acknowledge the hospitality of IPHT during their visits. We
thank A. Brown for his corrections.
\end{acknowledgments}

%-----------------------------------------------------------
\appendix%-----------------------------------------------------------

\section{Dressed-state Hamiltonian}

\subsection{Hamiltonian of the system}

Quantization of the resonator eigenmodes results in the following
expressions for the current and voltage operators and the Hamiltonian \cite%
{Om10}%
\begin{eqnarray}
I(x) &=&\sum \sqrt{\frac{\hbar \omega _{j}}{L_{\mathrm{r}}}}\left(
a_{j}+a_{j}^{\dag }\right) \cos k_{j}x, \\
V(x) &=&-i\sum \sqrt{\frac{\hbar \omega _{j}}{C_{\mathrm{r}}}}\left(
a_{j}-a_{j}^{\dag }\right) \sin k_{j}x, \\
H_{\mathrm{r}} &=&\sum \hbar \omega _{j}a_{j}^{\dag }a_{j},
\end{eqnarray}%
where $k_{j}=\pi j/l$, $\omega _{j}=vk_{j}$, $v=1/\sqrt{LC}$, and $a_{j}$
and $a_{j}^{\dag }$ are the annihilation and creation operators for the
photons of the $j$-th mode in the resonator. If the two signals (the
probing\ one and the driving one) are close to the fundamental and the third
harmonic components, we have the following for the relevant values: \newline
the current at $x=0$ (which defines the inductive coupling to the qubit, see
$H_{\mathrm{int}}$ below)%
\begin{equation}
I(0)=I_{\mathrm{r}0}[(a_{1}+a_{1}^{\dag })+\sqrt{3}(a_{3}+a_{3}^{\dag })],%
\text{ }I_{\mathrm{r}0}=\sqrt{\frac{\hbar \omega _{\mathrm{r}}}{L_{\mathrm{r}%
}}},
\end{equation}%
the voltage at the boundaries (which defines coupling the resonator to the
driving field, see $H_{\mathrm{\mu w}}$ below)%
\begin{equation}
V(\pm \frac{l}{2})=\pm iV_{\mathrm{r}0}[(a_{1}-a_{1}^{\dag })-\sqrt{3}%
(a_{3}-a_{3}^{\dag })],V_{\mathrm{r}0}=\sqrt{\frac{\hbar \omega _{\mathrm{r}}%
}{C_{\mathrm{r}}}},
\end{equation}%
and the resonator's Hamiltonian%
\begin{equation}
H_{\mathrm{r}}=\hbar \omega _{\mathrm{r}}a_{1}^{\dag }a_{1}+3\hbar \omega _{%
\mathrm{r}}a_{3}^{\dag }a_{3}.
\end{equation}

The interaction between the two-mode resonator and the flux qubit is
described with

\begin{eqnarray}
H_{\mathrm{int}} &=&MI(0)I_{\mathrm{qb}}=-\hbar \lbrack \mathrm{g}%
_{1}(a_{1}+a_{1}^{\dag })+\mathrm{g}_{3}(a_{3}+a_{3}^{\dag })]\tau _{z},%
\text{ \ \ }  \notag \\
\hbar \mathrm{g}_{1} &=&MI_{\mathrm{r}0}I_{\mathrm{p}},\text{ \ }\mathrm{g}%
_{3}=\sqrt{3}\mathrm{g}_{1},
\end{eqnarray}%
where $M$ is the mutual loop-resonator inductance.

The driving Hamiltonian originates from the energy of the left coupling
capacitance $C_{0}$, $H_{C_{0}}=C_{0}\Delta V^{2}$. Here the probe + drive
voltage to the left is the input voltage $V_{\mathrm{in}}=V_{\mathrm{p}}\sin
\omega _{\mathrm{p}}t+V_{\mathrm{d}}\sin \omega _{\mathrm{d}}t$, which
couples to the voltage in the resonator, $V(-l/2)$. Then leaving only the
slowly rotating terms, we obtain%
\begin{eqnarray}
H_{\mathrm{\mu w}} &=&\xi _{\mathrm{p}}(a_{1}e^{i\omega _{\mathrm{p}%
}t}+a_{1}^{\dag }e^{-i\omega _{\mathrm{p}}t})+\xi _{\mathrm{d}%
}(a_{3}e^{i\omega _{\mathrm{d}}t}+a_{3}^{\dag }e^{-i\omega _{\mathrm{d}}t}),
\notag \\
\text{\ }\xi _{\mathrm{p}} &=&\frac{1}{2}C_{0}V_{\mathrm{p}}V_{\mathrm{r}0}%
\text{, \ }\xi _{\mathrm{d}}=\frac{\sqrt{3}}{2}C_{0}V_{\mathrm{d}}V_{\mathrm{%
r}0}.
\end{eqnarray}

So, we have obtained the total Hamiltonian for the doubly-driven
qubit-resonator system (in which we rename the intracavity photon operators $%
a_{1}\equiv a$ and $a_{3}\equiv d$) :
\begin{eqnarray}
H_{\mathrm{tot}} &=&H_{\mathrm{qb}}+H_{\mathrm{r}}+H_{\mathrm{int}}+H_{%
\mathrm{\mu w}},  \label{H_qb-r} \\
H_{\mathrm{qb}} &=&-\frac{\Delta }{2}\tau _{x}-\frac{\varepsilon _{0}}{2}%
\tau _{z}, \\
H_{\mathrm{r}} &=&\hbar \omega _{\mathrm{r}}a^{\dag }a+3\hbar \omega _{%
\mathrm{r}}d^{\dag }d, \\
H_{\mathrm{int}} &=&-\hbar \lbrack \mathrm{g}_{1}(a+a^{\dag })+\mathrm{g}%
_{3}(d+d^{\dag })]\tau _{z},\text{ }  \label{Hint} \\
H_{\mathrm{\mu w}} &=&\xi _{\mathrm{p}}(ae^{i\omega _{\mathrm{p}}t}+a^{\dag
}e^{-i\omega _{\mathrm{p}}t})+ \\
&&+\xi _{\mathrm{d}}(de^{i\omega _{\mathrm{d}}t}+d^{\dag }e^{-i\omega _{%
\mathrm{d}}t}).  \notag
\end{eqnarray}

\subsection{Reduced Hamiltonian}

Thus, we have the Hamiltonian for the qubit - two-mode resonator system.
There, in respect to the experiment, we can assume the probe signal $\xi _{%
\mathrm{p}}$ to be relatively weak, so that $\left\langle n\right\rangle
=\left\langle a^{\dag }a\right\rangle \ll 1$, while the driving signal is
relatively strong, $\xi _{\mathrm{d}}\gg \xi _{\mathrm{p}}$, so that $%
\left\langle N\right\rangle =\left\langle d^{\dag }d\right\rangle \gg 1$.
Observables are related to the former mode, and hence the latter mode can be
traced out. In this way, one can obtain the reduced Hamiltonian, which
describes the system of the one-mode resonator + "dressed" qubit. Then one
can discuss the "dressed" relaxation as e.g. in Refs. [%
\onlinecite{Hauss08,
Wilson10}], which approach allows to proceed analytically.

We now consider the Hamiltonian of the qubit + third-harmonic driving in
order to transform it to the dressed-state qubit Hamiltonian,
\begin{eqnarray}
H_{\mathrm{qb+d.}} &=&-\frac{\Delta }{2}\tau _{x}-\frac{\varepsilon _{0}}{2}%
\tau _{z}+3\hbar \omega _{\mathrm{r}}d^{\dag }d+  \label{Hqb+d} \\
&&+\hbar \mathrm{g}_{3}\left( d+d^{\dag }\right) \tau _{z}+\xi _{\mathrm{d}%
}\left( de^{i\omega _{\mathrm{d}}t}+d^{\dag }e^{-i\omega _{\mathrm{d}%
}t}\right) .  \notag
\end{eqnarray}%
This does not include the probe signal (first harmonic), since all the
operations in this subsection would not affect it.

First, let us eliminate the time dependence from the driving by applying the
unitary operator $U_{1}=\exp (i\omega _{\mathrm{d}}td^{\dag }d)$. We obtain
the new Hamiltonian%
\begin{eqnarray}
H_{1} &=&U_{1}H_{\mathrm{qb+d.}}U_{1}^{\dag }+i\hbar \dot{U}_{1}U_{1}^{\dag
}= \\
&=&-\frac{\Delta }{2}\tau _{x}-\frac{\varepsilon _{0}}{2}\tau _{z}+\hbar
(3\omega _{\mathrm{r}}-\omega _{\mathrm{d}})d^{\dag }d+  \notag \\
&&+\hbar \mathrm{g}_{3}\left( de^{-i\omega _{\mathrm{d}}t}+d^{\dag
}e^{i\omega _{\mathrm{d}}t}\right) \tau _{z}+\xi _{\mathrm{d}}\left(
d+d^{\dag }\right) .  \notag
\end{eqnarray}%
For a large intracavity photon number $\left\langle N\right\rangle $, the
driving field can be described as coherent state. Averaging the Hamiltonian
with respect to the coherent state $\left\vert \alpha \right\rangle $ (note $%
d\left\vert \alpha \right\rangle =\alpha \left\vert \alpha \right\rangle $)
yields,%
\begin{equation}
H_{2}=\left\langle H_{1}\right\rangle =\left\langle \alpha \right\vert
H_{1}\left\vert \alpha \right\rangle =-\frac{\Delta }{2}\tau _{x}-\frac{%
\varepsilon _{0}+A_{\mathrm{d}}\cos \omega _{\mathrm{d}}t}{2}\tau _{z},
\end{equation}%
where $A_{\mathrm{d}}=4\alpha \hbar \mathrm{g}_{3}$, $\alpha =\sqrt{%
\left\langle N\right\rangle }$, and two constant terms $\hbar (3\omega _{%
\mathrm{r}}-\omega _{\mathrm{d}})\left\langle N\right\rangle $ and $2\alpha
\xi _{\mathrm{d}}$ have been omitted.

Next, we rewrite the Hamiltonian $H_{2}$ in the eigenbasis of its
time-independent part, using the transformation $S=\exp (i\varsigma \tau
_{y}/2)$ with $\tan \varsigma =-\Delta /\varepsilon _{0}$,%
\begin{equation}
H_{\mathrm{qb}}^{\prime }=-\frac{\Delta E}{2}\sigma _{z}-\frac{A_{\mathrm{d}%
}\cos \omega _{\mathrm{d}}t}{2}\left( \frac{\varepsilon _{0}}{\Delta E}%
\sigma _{z}-\frac{\Delta }{\Delta E}\sigma _{x}\right) ,  \label{H_in_eigen}
\end{equation}%
which allows for the subsequent application of the rotating-wave
approximation (RWA).

\subsection{RWA for weak driving}

For relatively weak driving ($A_{\mathrm{d}}<\hbar \omega _{\mathrm{d}}$) we
will use the conventional version of RWA. (Note that the conditions $A_{%
\mathrm{d}}<\hbar \omega _{\mathrm{d}}$ and $\left\langle N\right\rangle \gg
1$ are consistent for weak coupling $\mathrm{g}_{3}\ll \omega _{\mathrm{d}}$%
.) For this we make the transformation with $U=\exp (-i\omega _{\mathrm{d}%
}t\sigma _{z}/2)$, and omitting fast-rotating terms, obtain%
\begin{equation}
\widetilde{H}_{\mathrm{qb}}=UH_{\mathrm{qb}}^{\prime }U^{\dag }+i\hbar \dot{U%
}U^{\dag }\simeq -\frac{\widetilde{\varepsilon }}{2}\sigma _{z}+\frac{%
\widetilde{\Delta }}{2}\sigma _{x},  \label{Hqb}
\end{equation}%
where $\widetilde{\varepsilon }=\Delta E-\hbar \omega _{\mathrm{d}}$, $%
\widetilde{\Delta }=\Delta A_{\mathrm{d}}/2\Delta E$. Diagonalization of
this Hamiltonian gives the dressed energy levels, Eq.~(\ref{dressedDE}).
These transformations would also affect the interaction term, Eq.~(\ref{Hint}%
): $\tau _{z}\rightarrow \frac{\varepsilon _{0}}{\Delta E}\sigma _{z}-\frac{%
\Delta }{\Delta E}\sigma _{x}$, where the second term can be neglected in
RWA, while the first term results in the renormalization of the coupling
coefficient: $\mathrm{g}_{1}\rightarrow \mathrm{g}_{\varepsilon }=\mathrm{g}%
_{1}\varepsilon _{0}/\Delta E$. Collecting all together, we obtain
Hamiltonian (\ref{H_tilda}). Then we move to the dressed-qubit eigenstates,
using the transformation $\widetilde{S}=\exp (i\chi \sigma _{y}/2)$ with $%
\tan \chi =\widetilde{\Delta }/\widetilde{\varepsilon }$, and get
\begin{eqnarray}
\widetilde{H}^{\prime } &=&\frac{\widetilde{\Delta E}}{2}\widetilde{\sigma }%
_{z}+\hbar \omega _{\mathrm{r}}a^{\dag }a+\hbar \mathrm{g}_{\varepsilon
}\left( a+a^{\dag }\right) \times \\
&&\times \left( \frac{\widetilde{\varepsilon }}{\widetilde{\Delta E}}%
\widetilde{\sigma }_{z}+\frac{\widetilde{\Delta }}{\widetilde{\Delta E}}%
\widetilde{\sigma }_{x}\right) +\xi _{\mathrm{p}}\left( ae^{i\omega _{%
\mathrm{p}}t}+a^{\dag }e^{-i\omega _{\mathrm{p}}t}\right) .  \notag
\end{eqnarray}%
Then the second RWA is made with $U_{\mathrm{p}}=\exp \left[ i\omega _{%
\mathrm{p}}t\left( a^{\dag }a+\widetilde{\sigma }_{z}/2\right) \right] $,
and, omitting the fast-rotating terms, we obtain Hamiltonian (\ref{H_RWA}).

\subsection{RWA for strong driving}

To apply RWA to the Hamiltonian (\ref{H_in_eigen}) in the limit of strong
driving, we consider the following unitary transformation: $W=\exp (-i\eta
\sigma _{z}/2)$, where $\eta $ is given by the integral from the second term
in Eq.~(\ref{H_in_eigen}),
\begin{equation}
\eta =\frac{\varepsilon _{0}}{\Delta E}\frac{A_{\mathrm{d}}}{\hbar \omega _{%
\mathrm{d}}}\sin \omega _{\mathrm{d}}t.
\end{equation}%
This results in the new Hamiltonian%
\begin{eqnarray}
H_{\mathrm{qb}}^{\prime \prime } &=&-\frac{\Delta E}{2}\sigma _{z}+%
\widetilde{\Delta }\cos \omega _{\mathrm{d}}t\left( \sigma e^{i\eta }+\sigma
^{\dag }e^{-i\eta }\right) = \\
&=&-\frac{\Delta E}{2}\sigma _{z}+\frac{\widetilde{\Delta }}{2}\left(
e^{i\omega _{\mathrm{d}}t}+e^{-i\omega _{\mathrm{d}}t}\right) \times  \notag
\\
&&\times \sum\limits_{l=-\infty }^{\infty }J_{l}\left( \frac{A_{\mathrm{d}}}{%
\hbar \omega _{\mathrm{d}}}\frac{\varepsilon _{0}}{\Delta E}\right) \left(
\sigma e^{il\omega _{\mathrm{d}}t}+\sigma ^{\dag }e^{-il\omega _{\mathrm{d}%
}t}\right) .  \notag
\end{eqnarray}%
Then the RWA consists in applying another transformation, $W_{2}=\exp
(-ik\omega _{\mathrm{d}}t\sigma _{z}/2)$, and omitting the fast
time-dependent terms. We obtain in the vicinity of the $k$-th resonance,
where $\Delta E-k\hbar \omega _{\mathrm{d}}\ll \Delta E$, the following%
\begin{equation}
\widetilde{H}_{\mathrm{qb}}^{(k)}=-\frac{\widetilde{\varepsilon }^{(k)}}{2}%
\sigma _{z}+\frac{\widetilde{\Delta }^{(k)}}{2}\sigma _{x},  \label{H_qb_k}
\end{equation}%
where
\begin{eqnarray}
\widetilde{\varepsilon }^{(k)} &=&\Delta E-k\hbar \omega _{\mathrm{d}}, \\
\widetilde{\Delta }^{(k)} &=&\Delta \frac{k\hbar \omega _{\mathrm{d}}}{%
\varepsilon _{0}}J_{k}\left( \frac{A_{\mathrm{d}}}{\hbar \omega _{\mathrm{d}}%
}\frac{\varepsilon _{0}}{\Delta E}\right) .
\end{eqnarray}%
Then for the dressed energy distance we obtain $\widetilde{\Delta E}^{(k)}=%
\sqrt{\widetilde{\varepsilon }^{(k)2}+\widetilde{\Delta }^{(k)2}}$ or
explicitly \cite{Silveri13, Krech05}%
\begin{equation}
\widetilde{\Delta E}^{(k)}=\sqrt{\left( \Delta E-k\hbar \omega _{\mathrm{d}%
}\right) ^{2}+\left( \Delta \frac{k\hbar \omega _{\mathrm{d}}}{\varepsilon
_{0}}J_{k}\left( \frac{A_{\mathrm{d}}}{\hbar \omega _{\mathrm{d}}}\frac{%
\varepsilon _{0}}{\Delta E}\right) \right) ^{2}}.  \label{dressedDE_strong}
\end{equation}%
In particular, at weak driving only the levels with $k=1$ are relevant and $%
J_{1}(x)\approx x/2$; then $\widetilde{\Delta }^{(1)}\approx \widetilde{%
\Delta }$ and $\widetilde{\Delta E}^{(1)}$ coincides with $\widetilde{\Delta
E}$, obtained in the weak-driving limit, Eq.~(\ref{dressedDE}). Moreover, in
the limit $\varepsilon _{0}\gg \Delta $ this gives
\begin{equation}
\widetilde{\Delta E}^{(k)}=\sqrt{(\varepsilon _{0}-k\hbar \omega _{\mathrm{d}%
})^{2}+\left( \Delta J_{k}(A_{\mathrm{d}}/\hbar \omega _{\mathrm{d}})\right)
^{2}},
\end{equation}%
which is in agreement with the results of RWA used, e.g., in Refs. [%
\onlinecite{Wilson10, SAN12}].

Applying the above transformations would also affect the qubit-resonator
interaction term. As the result we obtain the total Hamiltonian in the RWA,
given by Eq.~(\ref{H_tilda}) with the substitutions: $\widetilde{\varepsilon
}\rightarrow \widetilde{\varepsilon }^{(k)}$ and $\widetilde{\Delta }%
\rightarrow \widetilde{\Delta }^{(k)}$.

%Solution in the dispersive consideration
%
%    Dressing of the qubit by strong driving can be described by renormalizing its energy splitting <cite>Wilson07</cite>:
%
%   ?>?_{k}(?_{d})=?J_{k}(?),  ?=4g?v(<n?>)/??_{d}=A_{d}/??_{d}.
%
%This gives the analytical formulas for the (multiphoton) excitation of the qubit. Essentially the same results can be obtained in the semiclassical approach <cite>Shevchenko10</cite>. There, the k-photon excitation appears close to the resonant parameters, given by the relation ??=k??_{d}. The upper-level occupation probability P_{up}(t) oscillates with the frequency ?_{R}=v((??-k??_{d})?+?_{k}?) with the renormalized splitting ?_{k}. The time-averaged probability is given by
%
%   P_{up}=(1/2)?((?_{k}?)/((??-k??_{d})?+?_{k}?)). <label>Pup</label>
%
%    Since qubit's energy difference differs from the frequency of the first mode, we can make use of the results presented in the Appendix of Ref.~<cite>Om10</cite> for the dispersive regime. The phase of the transmission coefficient is given by
%
%   tan?=(1/(2?))(((C_{r})/(C?)))?((L_{qb})/(L_{r})), <label>fi</label>
%
%where
%
%   L_{qb}=M?((?I_{qb})/(??)),  I_{qb}=I_{p}<?_{z}>. <label>Lqb</label>
%
%For the qubit in the ground state we obtain the dip of transmission phase
%
%   tan?=A[1+(??/?)?]^{-3/2},  A=(2/?)(((C_{r})/(C?)))?((?g??)/(?_{r}?)),
%
%while for the strongly driven "dressed" qubit we expect the response in the form of alternating peaks and dips described by Eqs.~<ref>Pup</ref>-<ref>Lqb</ref>. And this can be easily checked by comparing with the experiment...


\begin{thebibliography}{99}
\bibitem{Oli} W.D. Oliver, Y. Yu, J.C. Lee, K.K. Berggren, L.S. Levitov, and
T.P. Orlando, Science \textbf{310}, 1653 (2005).

\bibitem{Shnyr} V.I.~Shnyrkov, Th.~Wagner, D.~Born, S.N.~Shevchenko,
W.~Krech, A.N.~Omelyanchouk, E.~Il'ichev, and H.-G.~Meyer, Phys. Rev. B%
\textbf{\ 73}, 024506 (2006).

\bibitem{Wilson07} C.M. Wilson, T. Duty, F. Persson, M. Sandberg, G.
Johansson, and P. Delsing, Phys. Rev. Lett. \textbf{98}, 257003 (2007).

\bibitem{Mollow72} B.R. Mollow, Phys. Rev. A \textbf{5}, 2217 (1972).

\bibitem{Wu77} F.Y. Wu, S. Ezekiel, M. Ducloy, and B.R. Mollow, Phys. Rev.
Lett. \textbf{38}, 1077 (1977).

\bibitem{Zakrzewski91} J. Zakrzewski, M. Lewenstein, and T.W. Mossberg,
Phys. Rev. A \textbf{44}, 7717\ (1991).

\bibitem{Wallraff07} A. Wallraff, D.I. Schuster, A. Blais, J.M. Gambetta, J.
Schreier, L. Frunzio, M.H. Devoret, S.M. Girvin, and R.J. Schoelkopf, Phys.
Rev. Lett. \textbf{99}, 050501 (2007).

\bibitem{Papageorge12} A. Papageorge, A. Majumdar, E.D. Kim, and J. Vu\v{c}%
kovi\'{c}, New J. Phys. \textbf{14,} 013028 (2012).

\bibitem{Silveri13} M. Silveri, J. Tuorila, M. Kemppainen, and E. Thuneberg,
Phys. Rev. B \textbf{87}, 134505 (2013).

\bibitem{Reuther10} G.M. Reuther, D. Zueco, F. Deppe, E. Hoffmann, E.P.
Menzel, T. Wei\ss l, M. Mariantoni, S. Kohler, A. Marx, E. Solano, R. Gross,
and P. H\"{a}nggi, Phys. Rev. B \textbf{81}, 144510 (2010).

\bibitem{Zhao13} J. Zhao, Y. Yu, and B.-B. Jin, IEEE Trans. Appl. Supercond.
\textbf{23}, 1701705 (2013).

\bibitem{Bishop09} L.S. Bishop, J.M. Chow, J. Koch, A.A. Houck, M.H.
Devoret, E. Thuneberg, S.M. Girvin, and R. J. Schoelkopf, Nature Phys.
\textbf{5}, 105 (2009).

\bibitem{Fink09} J.M. Fink, M. Baur, R. Bianchetti, S. Filipp, M. G\"{o}ppl,
P.J. Leek, L. Steffen, A. Blais, and A. Wallraff, Phys. Scr. \textbf{T137},
014013 (2009).

\bibitem{Hocke12} F. Hocke, X. Zhou, A. Schliesser, T. Kippenberg, H. Huebl,
and R. Gross, New J. Phys. \textbf{14}, 123037 (2012).

\bibitem{Coh-Tan} C. Cohen-Tannoudji, J. Dupont-Rock, and G. Grynberg,
\textit{Atom-Photon Interactions. Basic Processes and Applications}, John
Wiley, New York (1998), Chap. 6.

\bibitem{Nakamura01} Y. Nakamura, Yu.A. Pashkin, and J.S. Tsai, Phys. Rev.
Lett. \textbf{87}, 246601 (2001).

\bibitem{Liu06} Y.X. Liu, C.P. Sun, and F. Nori, Phys. Rev. A \textbf{74},
052321 (2006).

\bibitem{Sun11} G. Sun, X. Wen, B. Mao, Y. Yu, J. Chen, W. Xu, L. Kang, P.
Wu, and S. Han, Phys. Rev. B \textbf{83}, 180507(R) (2011).

\bibitem{Yan01} M. Yan, E.G. Rickey, and Y. Zhu, Phys. Rev. A \textbf{64},
013412 (2001).

\bibitem{Alzar06} C.L. Garrido Alzar, H. Perrin, B.M. Garraway, and V.
Lorent, Phys. Rev. A \textbf{74}, 053413 (2006).

\bibitem{Saiko11} A.P. Saiko and R. Fedaruk, JETP Lett. \textbf{91}, 681
(2010).

\bibitem{Satanin13} A.M. Satanin, M.V. Denisenko, A.I. Gelman, and F. Nori,
arXiv:1305.4800 (2013).

\bibitem{Abovyan13} G.A. Abovyan and G.Yu. Kryuchkyan, Phys. Rev. A \textbf{%
88}, 033811 (2013).

\bibitem{Astafiev10} O.V. Astafiev, A.A. Abdumalikov, Jr., A.M. Zagoskin,
Yu.A. Pashkin, Y. Nakamura, and J.S. Tsai, Phys. Rev. Lett. \textbf{104},
183603 (2010).

\bibitem{Oelsner12} G. Oelsner, P. Macha, O.V. Astafiev, E. Il'ichev, M.
Grajcar, U. H\"{u}bner, B.I. Ivanov, P. Neilinger, and H.-G. Meyer, Phys.
Rev. Lett. \textbf{110}, 053602 (2013).

\bibitem{Koshino13} K. Koshino, H. Terai, K. Inomata, T. Yamamoto, W. Qiu,
Z. Wang, and Y. Nakamura, Phys. Rev. Lett. \textbf{110}, 263601 (2013).

\bibitem{green02} Ya.S.~Greenberg, A.~Izmalkov, M.~Grajcar, E.~Il'ichev,
W.~Krech, and H.-G.~Meyer, Phys. Rev. B\textbf{\ 66}, 224511 (2002).

\bibitem{Greenberg05} Ya.S. Greenberg, E. Il'ichev, and A. Izmalkov,
Europhys. Lett. \textbf{72}, 880 (2005).

\bibitem{Greenberg07} Ya.S. Greenberg, Phys. Rev. B \textbf{76}, 104520
(2007).

\bibitem{Hauss08} J. Hauss, A. Fedorov, S. Andr\'{e}, V. Brosco, C. Hutter,
R. Kothari, S. Yeshwanth, A. Shnirman, and G. Sch\"{o}n, New J. Phys.
\textbf{10}, 095018 (2008).

\bibitem{Mefed} A.E. Mefed and V.A. Atsarkin, Pis'ma Zh. Eksp. Teor. Fiz.
\textbf{25}, 233 (1977) [JETP Lett. \textbf{25}, 215 (1977)].

\bibitem{Oelsner10} G. Oelsner, S.H.W. van der Ploeg, P. Macha, U. H\"{u}%
bner, D. Born, S. Anders, E. Il'ichev, H.-G. Meyer, M. Grajcar, S. W\"{u}%
nsch, M. Siegel, A.N. Omelyanchouk, and O. Astafiev, Phys. Rev. B \textbf{81}%
, 172505 (2010).

\bibitem{Om10} A.N. Omelyanchouk, S.N. Shevchenko, Ya.S. Greenberg, O.
Astafiev and E. Il'ichev, Low Temp. Phys. \textbf{36}, 893 (2010).

\bibitem{Tuorila10} J. Tuorila, M. Silveri, M. Sillanp\"{a}\"{a}, E.
Thuneberg, Y. Makhlin, and P. Hakonen, Phys. Rev. Lett. \textbf{105}, 257003
(2010).

\bibitem{Greentree99} A.D. Greentree, C. Wei, and N.B. Manson, Phys. Rev. A
\textbf{59}, 4083 (1999).

\bibitem{ScullyZubairy} M.O. Scully and M.S. Zubairy, \textit{Quantum Optics}
(Cambridge, Cambridge University Press) (1997).

\bibitem{Ian10} H. Ian, Y.X. Liu, and F. Nori, Phys. Rev. A \textbf{81},
063823 (2010).

\bibitem{Wilson10} C.M. Wilson, G. Johansson, T. Duty, F. Persson, M.
Sandberg, and P. Delsing, Phys. Rev. B \textbf{81}, 024520 (2010).

\bibitem{Mu92} Y. Mu and C.M. Savage, Phys. Rev. A \textbf{46}, 5944 (1992).

\bibitem{Ashhab09} S. Ashhab, J.R. Johansson, A.M. Zagoskin, and F. Nori,
New J. Phys. \textbf{11}, 023030 (2009).

\bibitem{Ella13} L. Ella and E. Buks, arXiv:1210.6902.

\bibitem{SAN12} S.N. Shevchenko, S. Ashhab, and F. Nori, Phys. Rev. B
\textbf{85}, 094502 (2012).

\bibitem{Li12} J. Li, M.P. Silveri, K.S. Kumar, J.-M. Pirkkalainen, A. Veps%
\"{a}l\"{a}inen, W.C. Chien, J. Tuorila, M.A. Sillanp\"{a}\"{a}, P.J.
Hakonen, E.V. Thuneberg, and G. S. Paraoanu, Nat. Commun. \textbf{4}, 1420
(2013).

\bibitem{Krech05} W. Krech, D. Born, V. Shnyrkov, T. Wagner, M. Grajcar, E.
Il'ichev, H.-G. Meyer, and Y. Greenberg, IEEE Trans. Appl. Supercond.
\textbf{15}, 876 (2005).

\bibitem{com} It is necessary to distinguish the number of photons $n$ from
the average photon number $\langle n\rangle =\sum n\langle n|\rho |n\rangle $%
. The latter, while being small, contains contributions from many $n$ states.
\end{thebibliography}
\end{document}